\documentclass[floatfix,reprint,superscriptaddress,amsmath,amssymb,aps,prl]{revtex4-2}
\usepackage{stix2}
\usepackage{color}
\usepackage{graphicx}
\usepackage{dcolumn}
\usepackage{hyperref}
\usepackage{upgreek}
\hypersetup{colorlinks=true, citecolor=blue, urlcolor=blue, linkcolor=blue, filecolor=blue}
\usepackage{xr-hyper}
\begin{document}

\preprint{APS/123-QED}

\title{Manipulating Charge Distribution in Moir\'e Superlattices by Light}

\author{Ruiping Guo}         
\affiliation{State Key Laboratory of Low Dimensional Quantum Physics and
Department of Physics, Tsinghua University, Beijing, 100084, China}
\affiliation{Institute for Advanced Study, Tsinghua University, Beijing 100084, China}

\author{Haowei Chen}
\affiliation{State Key Laboratory of Low Dimensional Quantum Physics and
Department of Physics, Tsinghua University, Beijing, 100084, China}
\affiliation{New Cornerstone Science Laboratory, Department of Physics, University of Hong Kong, Hong Kong, China}
\affiliation{HKU-UCAS Joint Institute of Theoretical and Computational Physics at Hong Kong, Hong Kong, China}

\author{Wenhui Duan}
\email{duanw@tsinghua.edu.cn}
\affiliation{State Key Laboratory of Low Dimensional Quantum Physics and
Department of Physics, Tsinghua University, Beijing, 100084, China}
\affiliation{Institute for Advanced Study, Tsinghua University, Beijing 100084, China}
\affiliation{Frontier Science Center for Quantum Information, Beijing, China}

\author{Yong Xu}
\email{yongxu@mail.tsinghua.edu.cn}
\affiliation{State Key Laboratory of Low Dimensional Quantum Physics and
Department of Physics, Tsinghua University, Beijing, 100084, China}
\affiliation{Frontier Science Center for Quantum Information, Beijing, China}

\author{Chong Wang}
\email{chongwang@mail.tsinghua.edu.cn}
\affiliation{State Key Laboratory of Low Dimensional Quantum Physics and
Department of Physics, Tsinghua University, Beijing, 100084, China}
\affiliation{Frontier Science Center for Quantum Information, Beijing, China}

\date{\today}

\begin{abstract}
In ordinary solids, nonlinear optical responses are typically studied in terms of unit-cell averages due to the {\aa}ngstr\"om-scale lattice constants. In contrast, moir\'e superlattices, characterized by a large length scale, unlock an often-overlooked degree of freedom: intra-supercell spatial variations of local observables.
Here, we formulate the second-order direct current (DC) charge response in a spatially resolved manner, showing that even uniform optical illumination can drive a static, spatially non-uniform charge redistribution within a supercell. This effect is ubiquitous and cannot be forbidden by any crystalline symmetries.
Furthermore, we identify a dominant contribution arising from diverging analytical response coefficients, which leads to linear-in-time growth of the redistribution in the absence of relaxation. 
This growth is driven by the convergence or divergence of local DC photocurrents.
Applying our theory to twisted bilayer MoTe$_2$, we demonstrate strong, highly tunable charge modulation controlled by light intensity and frequency, opening a route to \textit{in situ, all-optical} control of moir\'e-periodic electrostatic potentials. Our work underscores the importance of intra-cell degrees of freedom, which enable a qualitatively richer class of nonlinear optical responses in moir\'e superlattices.
\end{abstract}

\maketitle

{\it Introduction.---}Two-dimensional (2D) moir\'e superlattices feature a small energy scale manifested in their flat minibands \cite{bistritzer2011moire,andrei2020graphene,balents2020superconductivity,andrei2021marvels,mak2022semiconductor,kennes2021moire}. By quenching kinetic energy and amplifying the effect of electron-electron interactions, these flat bands host a rich array of correlated and topological phases, including magnetism, superconductivity, charge-ordered states, and both integer and fractional quantum anomalous Hall insulators 
\cite{andrei2020graphene,balents2020superconductivity,andrei2021marvels,mak2022semiconductor,kennes2021moire,zhao2025exploring,cao2018unconventional,cao2018correlated,yankowitz2019tuning,lu2019superconductors,stepanov2020untying,sharpe2019emergent,xu2020correlated,huang2021correlated,jiang2019charge,serlin2020intrinsic,li2021quantum,cai2023signatures,park2023observation,zeng2023thermodynamic,xu2023observation,lu2024fractional,xie2024tunable,wang2024fractional,wang2025higher}.

The small energy scale follows from the large length scale of moir\'e superlattices. Despite rapid advances in flat band physics, the direct physical implications of this large moir\'e length scale remain underexplored.
In ordinary solids, the {\aa}ngstr\"om-scale lattice constants render spatial variations of physical quantities within a unit cell experimentally undetectable. 
In contrast, moir\'e lattice constants exceed those of the monolayers by orders of magnitude. Local observables --- such as charge density \cite{baeumer2015ferroelectrically,zhang2025plasmonic}, 
current density \cite{hu2023light,agarwal2024shift}, and polarization 
\cite{kim2024electrostatic,ding2024engineering,yasuda2021stacking,zhang2023visualizing,li2025unusual} --- can exhibit pronounced spatial inhomogeneities across the supercell, accessible to direct experimental probes. These inhomogeneities represent new degrees of freedom,  inviting systematic studies of their dynamics, manipulation methods, and coupling to external fields.


\begin{figure}[t]
\centering
\includegraphics[width = \columnwidth]{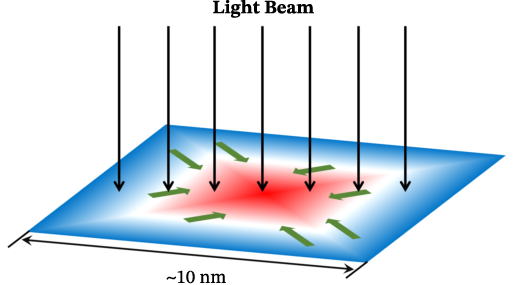}
\caption{Illustration of a moir\'e supercell under uniform optical illumination, showing in-plane, non-uniform DC photocurrents (green arrows) converging toward the cell center. Red and blue colors mark regions of positive and negative charge accumulation resulting from photocurrent convergence and divergence.}
\label{fig:concept}
\end{figure}

In this work, we investigate nonlinear optical responses of intra-cell degrees of freedom, focusing on the second-order DC intra-cell charge response. 
We show that uniform light at normal incidence induces static charge redistribution, leading to pronounced intra-cell charge variations. This effect is ubiquitous and cannot be forbidden by any crystalline symmetries.
Within the independent-particle approximation, we derive analytical expressions and identify a first-order divergence in the response coefficients, 
indicating that the charge density grows linearly in time in the absence of relaxation. 
This behavior stems from a non-zero divergence of the DC photocurrent.
To capture long-time saturation behavior, we introduce a phenomenological relaxation time $\uptau$, which determines the steady-state charge density. 
As an example, we apply our theory to twisted bilayer MoTe$_2$ (tMoTe$_2$), 
showing that the resulting charge pattern is highly tunable via light frequency. The response scales linearly with light intensity, enabling enhancement via stronger illumination. 
Our results reveal a mechanism 
to generate and control moir\'e-periodic electrostatic potentials by light, offering a platform for \textit{in-situ, all-optical} control of correlated phenomena in layered materials. Our work underscores the importance of intra-cell degrees of freedom, which enable a qualitatively richer class of nonlinear optical responses in moir\'e superlattices.

{\it General theory.---}We begin by demonstrating how the optically rectified photocurrents~\cite{sipe2000second,boyd2008nonlinear,chen2022basic} inevitably lead to non-uniform charge responses, highlighting their ubiquity and motivating a spatially resolved investigation. 
Consider a twisted van der Waals bilayer under spatially uniform illumination. Even though the driving electric field is homogeneous, the resulting DC photocurrent exhibits spatial modulation across the moir\'e supercell.
Crucially, no symmetry or physical constraint enforces a divergence-free DC photocurrent. 
By the continuity equation $\partial_t \rho + \nabla \cdot \mathbf{j} = 0$, where $\rho$ is the charge density and $\mathbf{j}$ is the photocurrent density, 
any non-zero divergence leads to local charge accumulation over time (see Fig.~\hyperref[fig:concept]{1}). This mechanism explains how uniform light can generate inhomogeneous charge distributions.

The above analysis applies in the intermediate regime --- longer than the optical cycle but shorter than the relaxation time $\uptau$ --- before reaching steady state. 
In this regime, the transient~\cite{belinicher1982kinetic,sturman2020ballistic,alexandradinata2023anomalous} DC photocurrent can exhibit non-zero divergence, driving charge inhomogeneity. 
On longer timescales, charge accumulation saturates at a level determined by $\uptau$. Thus, charge inhomogeneity can be exceptionally large in high-quality samples with large $\uptau$.

\begin{figure}[b]
    \centering
\includegraphics[width = \columnwidth]{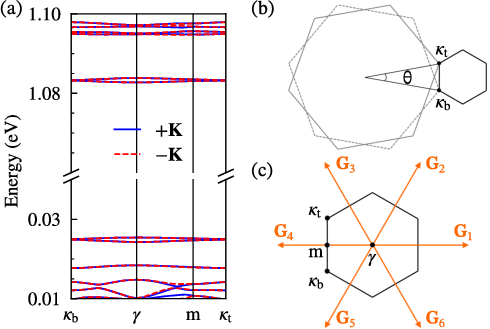}
\caption{(a) Band structure of $\text{tMoTe}_{\text{2}}$ at twist angle $\theta=1.0^\circ$. The $\pm \mathbf{K}$ valley degrees of 
freedom, inherited from monolayer $\text{MoTe}_{\text{2}}$, remain decoupled. Blue solid and red dashed lines are bands from the $+\mathbf{K}$ and $-\mathbf{K}$ valleys, respectively. 
The band gap is $1.057$~eV.
(b) Brillouin zones of the top (solid gray), bottom (dashed gray) monolayer MoTe$_2$, and the moir\'e Brillouin zone of tMoTe$_2$ (black). 
(c) Enlarged moir\'e Brillouin zone, showing high-symmetry points and six reciprocal lattice vectors ($\mathbf{G}_{1-6}$) with the smallest magnitude.}
\label{fig:tMoTe2}
\end{figure}

Consider 2D moir\'e superlattices illuminated by normally incident monochromatic light of frequency $\omega_0$. The 
electric field can be written as $\mathbf{E}(t) = \mathbf{E}_0(\omega_0) \, e^{-i \omega_0 t} + \mathbf{E}_0(-\omega_0) \, e^{i \omega_0 t}$, where $\mathbf{E}_0(-\omega_0) = \mathbf{E}^{*}_0(\omega_0)$. Without illumination, the charge density is denoted as $\rho_{\text{dark}}(\mathbf{r})$; under illumination it changes by $\Delta\rho(\mathbf{r})$. At second order in $\mathbf{E}$, the static response of $\Delta\rho(\mathbf{r})$ is
\begin{equation}
\label{eq:definition}
    \Delta\rho_{\mathbf{G}} = \zeta^{\alpha_1 \alpha_2}(\mathbf{G}; \omega_0) \; \mathrm{E}^{\alpha_1}_0(\omega_0)
    \mathrm{E}^{\alpha_2}_0(-\omega_0),
\end{equation}
where $\Delta\rho_{\mathbf{G}}$ is the Fourier 
component of $\Delta\rho(\mathbf{r})$, defined as $\Delta\rho(\mathbf{r}) = \sum_{\mathbf{G}} \; \Delta\rho_{\mathbf{G}} \, e^{i \mathbf{G} \cdot \mathbf{r}}$. Here, $\mathbf{G}$ is a moir\'e reciprocal lattice vector, and responses at wavevectors $\mathbf{q}\ne\mathbf{G}$ vanish due to discrete translational symmetry. In addition, the uniform component $\Delta\rho_{\mathbf{G}=\mathbf{0}}$ vanishes by charge conservation. In Eq.~(\ref{eq:definition}), $\alpha_1,\alpha_2 \in \{x,y\}$, and repeated indices imply summation. The response coefficient $\zeta^{\alpha_1 \alpha_2}(\mathbf{G}; \omega_0)$ 
is analogous to the second-order conductivity tensor $\sigma^{\beta \alpha_1 \alpha_2}$ in photocurrent responses~\cite{sipe2000second,boyd2008nonlinear,chen2022basic}.

In nonlinear optics, second-order responses in centrosymmetric systems are commonly regarded as vanishing~\cite{baltz1981theory,boyd2008nonlinear}. However, this applies only to spatially averaged quantities, such as polarization or current averaged over the unit cell. 
In contrast, we are interested in spatially resolved charge modulations within the moir\'e supercell, which are not symmetry-forbidden.
Even with inversion symmetry, the only constraint on $\zeta^{\alpha_1 \alpha_2}(\mathbf{G}; \omega_0)$ is 
$\zeta^{\alpha_1 \alpha_2}(\mathbf{G}; \omega_0) = \zeta^{\alpha_1 \alpha_2}(-\mathbf{G}; \omega_0)$, permitting non-zero responses. 
More generally, any point-group symmetry operation $\hat{g}$ imposes
\begin{equation}
    \zeta^{\alpha_1 \alpha_2}(\hat{g} \, \mathbf{G}; \omega_0) = g^{\alpha_1 \beta_1} \;
    \zeta^{\beta_1 \beta_2}(\mathbf{G}; \omega_0) \; g^{\alpha_2 \beta_2},
\end{equation}
where $\hat{g}$ acts on vectors as $v^{\alpha} \to g^{\alpha \beta} v^{\beta}$. These symmetry relations constrain the spatial structure of the responses without forcing it to vanish.

Within the independent-particle approximation~\footnote{The formation of a spatially modulated charge distribution under uniform illumination is rooted in the continuity equation and is therefore fully general, persisting even in the presence of excitonic effects and other interactions beyond the independent-particle approximation.}, $\zeta^{\alpha_1 \alpha_2}(\mathbf{G}; \omega_0)$ is derived analytically via perturbation theory. 
The optical field enters through the length gauge \cite{peres2017,peres2018}, as a dipole term $e\,\mathbf{E}(t)\cdot\mathbf{r}$, 
with the elementary charge $e>0$. Detailed steps are given in the Supplemental Material (SM) \cite{supp}\nocite{bradlyn2024uniformity}, Sec.~\ref{sec:derivation}. 
The final result is
\begin{eqnarray}
\label{eq:zeta}
    \zeta^{\alpha_1 \alpha_2}(\mathbf{G}; \omega_0) &&= e^3 \int \left[ \mathrm{d}\mathbf{k} \right] \sum_{s,s^{\prime}} 
    F^{\mathbf{G}}_{\mathbf{k} s^{\prime} s} \frac{1}{2i\hbar\eta-\Delta\epsilon_{\mathbf{k} s s^{\prime}}} \nonumber\\
    &&\times\left[ \mathrm{D}^{\alpha_2}_{\mathbf{k}},\frac{1}{\hbar\omega_0-\Delta\epsilon_{\mathbf{k}}+i\hbar\eta} 
    \circ[\mathrm{D}^{\alpha_1}_{\mathbf{k}}, \mathcal{P}^{[0]}_{\mathbf{k}}] \right]_{s s^{\prime}} \nonumber\\
    &&+ (\alpha_1,\omega_0 \leftrightarrow \alpha_2,-\omega_0),
\end{eqnarray}
where $\int \left[ \mathrm{d}\mathbf{k} \right] \equiv \int_{\text{MBZ}} \frac{\mathrm{d}\mathbf{k}}{(2\pi)^2}$ denotes moir\'e Brillouin zone integration, 
and the summation runs over bands $s,\,s^{\prime}$.
$\Delta\epsilon_{\mathbf{k} s s^{\prime}} \equiv \epsilon_{\mathbf{k} s} - \epsilon_{\mathbf{k} s^{\prime}}$ is the band energy difference, and $\eta=0^+$ denotes positive infinitesimal. 
$F^{\mathbf{G}}_{\mathbf{k} s s^{\prime}} \equiv \left\langle u_{\mathbf{k} s} \middle| u_{\mathbf{k}+\mathbf{G} s^{\prime}} \right\rangle$ is 
the overlap between periodic parts of Bloch states. $\mathrm{D}^{\alpha}_{\mathbf{k} s s^{\prime}} \equiv \delta_{s s^{\prime}} \nabla^{\alpha}_{\mathbf{k}} - 
i \xi^{\alpha}_{\mathbf{k} s s^{\prime}}$ is the covariant derivative \cite{peres2017,peres2018}, with Berry connection
$\xi^{\alpha}_{\mathbf{k} s s^{\prime}} \equiv i \left\langle u_{\mathbf{k} s} \right| \nabla^{\alpha}_{\mathbf{k}} \left| u_{\mathbf{k} s^{\prime}} \right\rangle$ . $\mathcal{P}^{[0]}_{\mathbf{k} s s^{\prime}} \equiv f_{\mathbf{k} s} \delta_{s s^{\prime}}$, with $f_{\mathbf{k} s}$ the Fermi-Dirac distribution. 
We use compact matrix notations for band indices: $[A,B] \equiv AB-BA$ (commutator), and 
$(A\circ B)_{s s^{\prime}} \equiv A_{s s^{\prime}} B_{s s^{\prime}}$ (Hadamard product). 

\begin{figure}[t]
\centering
\includegraphics[width = \columnwidth]{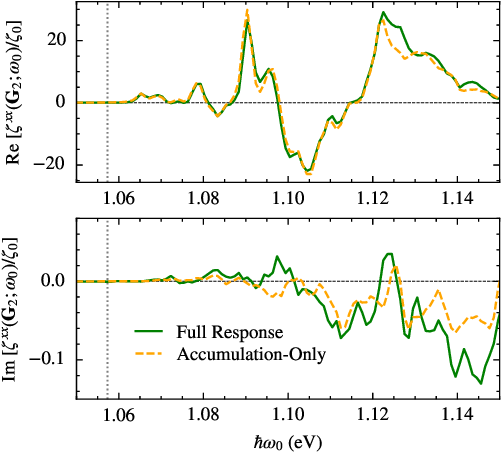}
\caption{Frequency dependence of $\zeta^{xx}(\mathbf{G}_2; \omega_0)$. Real and imaginary parts are shown in the top and bottom subplots, respectively. Here, $\zeta_0 \approx 1.602\times 10^{-19} \, \mathrm{C}^3/\mathrm{J}^2$. 
The full charge response [Eq.~(\ref{eq:zeta})] is shown as solid green lines; the accumulation-only contribution 
[Eq.~(\ref{eq:zeta_inj})] is indicated by dashed orange lines. Vertical dotted lines mark the band gap energy at 
$\hbar \omega_0 \approx 1.057~\text{eV}$}
\label{fig:frequency_dependence}
\end{figure}

In Eq.~(\ref{eq:zeta}), terms with $s=s^{\prime}$ yield an $\mathcal{O}(\eta^{-1})$ divergence, isolated as (see SM \cite{supp}, Sec.~\ref{sec:derivation})
\begin{eqnarray}
\label{eq:zeta_inj}
    \zeta_{\mathrm{accum}}^{\alpha_1 \alpha_2}(\mathbf{G}; \omega_0) =&& \frac{\pi e^3}{\hbar\eta} 
    \int \left[ \mathrm{d}\mathbf{k} \right] \sum_{r \ne s} \left( F^{\mathbf{G}}_{\mathbf{k} ss} - F^{\mathbf{G}}_{\mathbf{k} rr} \right) \nonumber\\
    \times&& \xi^{\alpha_1}_{\mathbf{k} rs} \xi^{\alpha_2}_{\mathbf{k} sr} \Delta f_{\mathbf{k} sr} \delta(\hbar\omega_0 - \Delta\epsilon_{\mathbf{k} rs}),
\end{eqnarray}
where $\Delta f_{\mathbf{k} sr} \equiv f_{\mathbf{k} s} - f_{\mathbf{k} r}$. The subscript ``accum'' denotes photocurrent accumulation, as justified below. 
Equation~(\ref{eq:zeta}) is general, but here Eq.~(\ref{eq:zeta_inj}) assumes a semiconductor at zero temperature for simplicity. 
Such divergences in nonlinear optical response coefficients are well known \cite{fregoso2018jerk,fregoso2019bulk,ventura2020second,gao2021intrinsic}. 
For instance, injection currents \cite{sipe2000second,chan2017photocurrents,de2017quantized,zhang2019switchable,gao2021intrinsic} grow linearly in time in the absence of relaxation, corresponding to an $\mathcal{O}(\eta^{-1})$ divergence in the second-order optical conductivity $\sigma^{\beta \alpha_1 \alpha_2}$. 
Jerk currents \cite{fregoso2018jerk,fregoso2019bulk,ventura2020second}, with constant second-order time derivatives and $\mathcal{O}(\eta^{-2})$ divergences in $\sigma^{\beta \alpha_1 \alpha_2 \alpha_3}$, constitute another example. 
Similar to the discussion in Refs.~\cite{fregoso2018jerk,fregoso2019bulk,ventura2020second,xie2025photon,gao2021intrinsic}, such divergences imply a finite, constant time derivative of the charge response:
\begin{equation}
\label{eq:linear_growth}
    \frac{\mathrm{d}\Delta\rho_{\mathbf{G}}}{\mathrm{d} t} = 2\eta\, \zeta_{\mathrm{accum}}^{\alpha_1 \alpha_2}(\mathbf{G}; \omega_0) \,
    \mathrm{E}_0^{\alpha_1}(\omega_0) \, \mathrm{E}_0^{\alpha_2}(-\omega_0).
\end{equation}
Equation~(\ref{eq:linear_growth}) holds only without relaxation---i.e., in the time window $1/\omega_0 \ll t \ll \uptau$. 
For $t \gtrsim \uptau$, $\Delta\rho_{\mathbf{G}}$ saturates to a value determined by $\uptau$. 
Introducing relaxation processes phenomenologically by replacing $\eta\rightarrow1/\uptau$ \cite{peres2018,ventura2020second,zhang2019switchable,gao2021intrinsic}, which is 
formally equivalent to switching on the external field adiabatically, the saturated charge density is given by 
$\Delta\rho_{\mathbf{G}} = \frac{\uptau}{2} \frac{\mathrm{d}\Delta\rho_{\mathbf{G}}}{\mathrm{d} t}$, 
which is proportional to both $\uptau$ and the time derivative in Eq.~(\ref{eq:linear_growth}). In high-quality samples with large $\uptau$,  
$\zeta_{\mathrm{accum}}^{\alpha_1 \alpha_2}(\mathbf{G}; \omega_0)$ dominates the total response $\zeta^{\alpha_1 \alpha_2}(\mathbf{G}; \omega_0)$. 
We illustrate this by numerically evaluating the full and accumulation-only charge responses for the tMoTe$_2$ model, whose electronic structure is shown in Fig.~\hyperref[fig:tMoTe2]{2}; 
the resulting curves are shown in Fig.~\hyperref[fig:frequency_dependence]{3}, with a detailed discussion provided later.

\begin{figure*}
    \centering
\includegraphics[width = \textwidth]{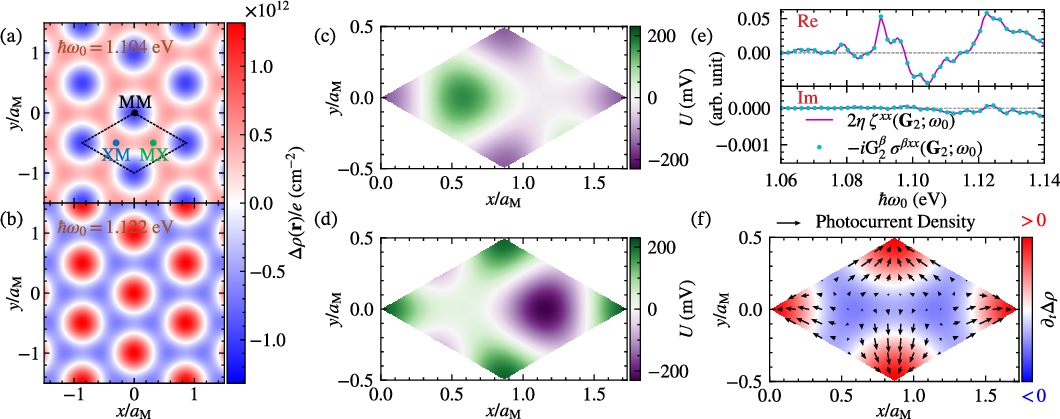}
\caption{Real-space charge redistribution in $\text{tMoTe}_{\text{2}}$ under linearly polarized illumination.
(a,b) Photoinduced charge density $\Delta\rho(\mathbf{r})$ under $x$-polarized light at photon energies $\hbar \omega_0 = 1.104 \, \mathrm{eV}$ and 
$1.122 \, \mathrm{eV}$, respectively, with fixed intensity $1.0\times 10^{11} \, \mathrm{W/m^2}$. The dashed parallelogram in (a) outlines a moir\'e supercell, with high-symmetry stacking sites MM, XM, and MX marked by black, blue, and green dots. 
(c,d) Moir\'e-periodic electrostatic potentials $U(\mathbf{r})$ induced in a nearby layered material, computed from the charge distributions in (a,b), respectively. (e) Comparison between the charge response coefficient $2\eta\,\zeta^{xx}(\mathbf{G}_2; \omega_0)$ computed directly (solid pink) and inferred from the continuity equation [Eq.~(\ref{eq:CE_coef})] 
via the corresponding photocurrent (cyan dots). 
(f) Spatial distribution of the DC photocurrent (black arrows) and time derivative $\partial_t \Delta\rho$ (color map) under the same conditions as in (b), within a single moir\'e supercell.}
\label{fig:patterns}
\end{figure*}

Before concluding this section, we provide a real-space interpretation of the constant time derivative of $\Delta\rho_{\mathbf{G}}$ in 
Eq.~(\ref{eq:linear_growth}): it originates from the non-zero divergence of the non-uniform DC photocurrent, which acts as a local charge source. 
By local charge conservation, the charge and current responses are related via the continuity equation, 
$\partial_t \Delta\rho + \nabla \cdot \mathbf{j} = 0$. Defining $\mathbf{j}_{\mathbf{G}}$ as the Fourier component of the DC photocurrent 
$\mathbf{j}(\mathbf{r}) = \sum_{\mathbf{G}} \; \mathbf{j}_{\mathbf{G}} \, e^{i \mathbf{G} \cdot \mathbf{r}}$, the equation in momentum space becomes:
\begin{equation}
\label{eq:continuity_equation}
    \frac{\mathrm{d}\Delta\rho_{\mathbf{G}}}{\mathrm{d} t} + i \, \mathbf{G} \cdot \mathbf{j}_{\mathbf{G}} = 0,
\end{equation}
where $\mathbf{G}$ runs over moir\'e reciprocal lattice vectors. According to Eq.~(\ref{eq:continuity_equation}), if 
$i \, \mathbf{G} \cdot \mathbf{j}_{\mathbf{G}} \neq 0$ for a time-independent $\mathbf{j}$ (as in shift currents), then $\Delta\rho_{\mathbf{G}}$ 
grows linearly in time. 
This supports the view that photocurrent divergence continuously injects charge. 
To verify Eq.~(\ref{eq:continuity_equation}) in our framework, we compute both terms independently. The time derivative 
$\frac{\mathrm{d}\Delta\rho_{\mathbf{G}}}{\mathrm{d} t}$ follows from Eq.~(\ref{eq:linear_growth}); 
the current term $i \, \mathbf{G} \cdot \mathbf{j}_{\mathbf{G}}$ is obtained via perturbation theory within the independent-particle approximation.
The non-uniform photocurrent is given by 
$\mathrm{j}^{\beta}_{\mathbf{G}} = \sigma^{\beta \alpha_1\alpha_2}(\mathbf{G};\omega_0) \mathrm{E}^{\alpha_1}_0(\omega_0) \mathrm{E}^{\alpha_2}_0(-\omega_0)$, 
analogous to the charge response in Eq.~(\ref{eq:definition}). The full expression for $\sigma^{\beta \alpha_1\alpha_2}(\mathbf{G};\omega_0)$ 
appears in Sec.~\ref{sec:derivation} of the SM \cite{supp}. In terms of response coefficients, Eq.~(\ref{eq:continuity_equation}) becomes
\footnote{Equation~(\ref{eq:CE_coef}) remains valid even when $\eta$ is finite (i.e., not taken to be infinitesimal), provided that 
$\zeta^{\alpha_1\alpha_2}(\mathbf{G};\omega_0)$ is interpreted as the full charge response coefficient given by Eq.~(\ref{eq:zeta}), 
rather than the accumulation-only contribution in Eq.~(\ref{eq:zeta_inj}); see SM \cite{supp}, Sec.~\ref{sec:time_derivative} for a derivation. In the limit 
$\eta \to 0^+$, only $\zeta_{\text{accum}}^{\alpha_1\alpha_2}(\mathbf{G};\omega_0)$ survives. However, for numerical computations, a small but finite value 
of $\eta$ must be used in the expression of $\sigma^{\beta \alpha_1\alpha_2}(\mathbf{G};\omega_0)$, making the generalized form of Eq.~(\ref{eq:CE_coef}) 
discussed here useful.}:
\begin{equation}
\label{eq:CE_coef}
    2\eta\,\zeta^{\alpha_1\alpha_2}(\mathbf{G};\omega_0) + i \mathrm{G}^{\beta} \sigma^{\beta \alpha_1\alpha_2}(\mathbf{G};\omega_0) = 0. 
\end{equation}
This relation holds with high numerical accuracy in our $\text{tMoTe}_{\text{2}}$ calculations; see Figs.~\hyperref[fig:patterns]{4(e)} and~\hyperref[fig:patterns]{4(f)} 
and related discussions. We thus confirm that charge redistribution is driven by photocurrent divergence.

{\it Twisted bilayer $\text{MoTe}_{\text{2}}$.---}Among several moir\'e materials we examined, tMoTe$_2$ exhibits the strongest optical response within our framework. We therefore numerically evaluate Eqs.~(\ref{eq:zeta}) and~(\ref{eq:zeta_inj}) for tMoTe$_2$ at a twist angle of $\theta=1.0^\circ$, 
where the top layer is rotated counterclockwise relative to the bottom one. Our calculations use a four-band continuum model incorporating both conduction and valence bands, 
following \cite{wu2019topological} (see SM \cite{supp}, Sec.~\ref{sec:continuum_model} for details). The corresponding band structure and moir\'e 
Brillouin zone are shown in Fig.~\ref{fig:tMoTe2}. 
This moir\'e electronic structure --- and the resulting charge response --- arises from interlayer interactions, not from a mere geometric superposition of two isolated monolayers.
We find that $\zeta^{\alpha_1 \alpha_2}(\mathbf{G}; \omega_0)$ decays rapidly with increasing $|\mathbf{G}|$, becoming negligible beyond the six smallest moir\'e 
reciprocal lattice vectors [orange arrows in Fig.~\hyperref[fig:tMoTe2]{2 (c)}].

To illustrate the charge response, we compute $\zeta^{\alpha_1 \alpha_2}(\mathbf{G}; \omega_0)$ for various 
$\alpha_1$, $\alpha_2$ and $\mathbf{G}$ using Eqs.~(\ref{eq:zeta}) and~(\ref{eq:zeta_inj}), and present a representative component, 
$\zeta^{xx}(\mathbf{G}_2; \omega_0)$ in Fig.~\ref{fig:frequency_dependence}. 
We take $\hbar \eta = 1\,\mathrm{meV}$, smaller than all other energy scales in the model and consistent with reported values for similar materials \cite{zhang2019switchable,kozawa2014photocarrier,sohier2018mobility}. 
As shown in Fig.~\ref{fig:frequency_dependence}, the response vanishes 
below the band gap and exhibits sharp peaks and oscillations above it. Notably, the 
accumulation contribution $\zeta^{xx}_{\mathrm{accum}}(\mathbf{G}_2; \omega_0)$ closely follows the full response, indicating its dominance under these parameters.

With the charge response coefficient $\zeta^{\alpha_1 \alpha_2}(\mathbf{G}; \omega_0)$ determined, we compute the real-space charge density response 
$\Delta\rho(\mathbf{r})$ using Eq.~(\ref{eq:definition}). The incident light intensity 
\footnote{In principle, both pulsed and continuous excitation can be employed experimentally. If pulsed excitation is used, the pulse duration must be sufficiently long for the steady-state charge redistribution to be established.} 
is set to $1.0 \times 10^{11}\,\mathrm{W/m^2}$, 
consistent with typical experiments \cite{kim2014ultrafast}, and we systematically vary the photon energy to explore its effects. Figures~\hyperref[fig:patterns]{4(a)} 
and~\hyperref[fig:patterns]{4(b)} display representative charge redistribution patterns under $x$-polarized illumination. 
At $\hbar \omega_0 = 1.122 \, \mathrm{eV}$, the peak value of $\Delta\rho(\mathbf{r})/e$ reaches  $1.3\times10^{12}\,\mathrm{cm}^{-2}$. 
Remarkably, the sign of the charge density reverses between $\hbar \omega_0 = 1.104 \, \mathrm{eV}$ and $\hbar \omega_0 = 1.122 \, \mathrm{eV}$, while its magnitude remains large --- demonstrating strong frequency tunability.

{\it Real-space interpretation.---}Within the continuum model of $\text{tMoTe}_{\text{2}}$, we compute the non-uniform DC photocurrent $\mathbf{j}(\mathbf{r})$ using 
the method detailed in the SM~\cite{supp}, Sec.~\ref{sec:derivation}, enabling direct analysis of the real-space origin of charge redistribution. 
Figure~\hyperref[fig:patterns]{4(e)} compares $2\eta\,\zeta^{xx}(\mathbf{G}_2;\omega_0)$ computed 
directly from Eq.~(\ref{eq:zeta}) and indirectly via the continuity equation, Eq.~(\ref{eq:CE_coef}). The excellent agreement across all frequencies confirms 
the validity of the continuity equation. Figure~\hyperref[fig:patterns]{4(f)} shows $\partial_t\Delta\rho$ 
under the same illumination conditions as in Fig.~\hyperref[fig:patterns]{4(b)}, closely tracking the divergence of the photocurrent: 
regions of current convergence (divergence) correspond to charge accumulation (depletion). 
These results confirm that the charge response mainly arises from the divergence of the non-uniform photocurrent.

{\it Applications.---}A promising application of the light-induced charge redistribution is the creation of an \textit{all-optical, in-situ} tunable moir\'e potential in a 
nearby 2D material. This potential arises from the electrostatic field generated by the photoinduced charge inhomogeneity. 
For instance, an additional 2D layer such as graphene or a transition metal dichalcogenide monolayer can be placed atop the $\text{tMoTe}_{\text{2}}$. 
Without illumination, the moir\'e modulation from twisting induces a charge distribution $\rho_{\text{dark}}(\mathbf{r})$ in $\text{tMoTe}_{\text{2}}$. 
Upon illumination at appropriate photon energies, additional charge redistribution generates an electrostatic potential $U(\mathbf{r})$ 
comparable to --- or even exceeding --- that from $\rho_{\text{dark}}$, enabling dynamic optical control. 
Assuming a $10$ \AA~vertical separation and using the same illumination conditions as in 
Figs.~\hyperref[fig:patterns]{4(a)}-\hyperref[fig:patterns] {4(b)}, we compute $U(\mathbf{r})$ by solving the Poisson equation (see SM \cite{supp}, Sec.~\ref{sec:poisson}); 
this separation avoids strong interlayer coupling --- otherwise the layers would need to be treated as a single electronic system --- 
and the results in Figs.~\hyperref[fig:patterns]{4(c)}-\hyperref[fig:patterns]{4(d)} are qualitatively insensitive to the precise separation. 
The peak value of $U(\mathbf{r})$ reaches $\sim200\,\text{mV}$ and can be further enhanced by increasing light intensity or using samples with higher quality (larger $\uptau$). 
The potential also exhibits strong tunability with photon energy. Such optically controlled moir\'e potentials may enable topological band engineering 
\cite{tan2024designing,zhan2025designing,ma2025magnetic,aultmccoy2025optimizing,liang2025generic}, 
exciton manipulation \cite{cho2024moir,kiper2024confined,zhao2021universal,kim2024electrostatic,zhang2025engineering,gu2025quantum}, 
transport control \cite{wang2024band,sequeira2024manipulating}, and enhanced optoelectronic responses \cite{nabil2025giant}.

Beyond influencing nearby 2D materials, the photoinduced charge inhomogeneity may be directly probed using real-space microscopic techniques 
\cite{nuckolls2024microscopic}. 
Potential methods include scanning tunneling microscopy \cite{li2024quasiperiodic,li2024tuning,woods2014commensurate,deng2020interlayer,chiu2025high}, 
Kelvin probe force microscopy \cite{han2024highly}, 
atomic single-electron transistors \cite{klein2024imaging}, and transmission electron microscopy \cite{mishima2024atomic}, all capable of resolving charge distributions at the moir\'e scale.

{\it Conclusions.---}We introduce a general mechanism for light‐driven charge redistribution in moir\'e superlattices.
In systems with large $\uptau$, it primarily stems from the spatial convergence and divergence of non-uniform DC photocurrents. 
Since the response scales linearly with illumination intensity, it can be readily enhanced by stronger light intensity in experiments.
This mechanism enables an \textit{all-optical, in-situ} tunable moir\'e potential in adjacent layered materials, offering contact-free means to control electronic properties.
Looking ahead, it will be important to explore how this phenomenon intertwines with band‐geometry effects \cite{tan2016enhance,morimoto2016topological,nagaosa2017concept,xie2025photon}, to study its interplay with charge and spin ordering, to quantify the influence of many‐body interactions and lattice relaxation, and to develop experimental strategies for directly observing and harnessing light‐induced moir\'e potentials.

\begin{acknowledgments}
{\it Acknowledgements.---}We acknowledge helpful discussions
with Chen Hu, Dezhao Wu, and Biao Lian. This work
was supported by the National Key Basic Research
and Development Program of China (Grant
No. 2024YFA1409100), the Basic Science Center Project
of NSFC (Grant No. 52388201), the National Natural
Science Foundation of China (Grants No. 12574269,
No. 12334003, No. 12421004, and No. 12361141826),
the National Key Basic Research and Development
Program of China (Grant No. 2023YFA1406400), the
National Science Fund for Distinguished Young Scholars
(Grant No. 12025405), the Fundamental and
Interdisciplinary Disciplines Breakthrough Plan of the
Ministry of Education of China (Grant
No. JYB2025XDXM408), the Innovation Program for
Quantum Science and Technology (Grant
No. 2023ZD0300500) and Beijing Key Laboratory of
Quantum AI. The calculations were performed at National
Supercomputer Center in Tianjin using the Tianhe new
generation supercomputer
\end{acknowledgments}

\bibliography{ref}

\end{document}


\onecolumngrid
\vspace{1cm}
\begin{center}
{\bf\large Supplemental Material for\\$ $\\``Manipulating Charge Distribution in Moir\'e Superlattices by Light"}
\end{center}

\setcounter{secnumdepth}{3}
\setcounter{equation}{0}
\setcounter{figure}{0}
\renewcommand{\theequation}{S\arabic{equation}}
\renewcommand{\thefigure}{S\arabic{figure}}
\renewcommand\figurename{Supplementary Figure}
\renewcommand\tablename{Supplementary Table}
\newcommand\Scite[1]{[{\color{blue}S}\citealp{#1}]}
\makeatletter \renewcommand\@biblabel[1]{[S#1]} \makeatother

This supplemental material provides additional technical details supporting the main text. It is organized as follows:

\begin{itemize}
\item Sec.~\ref{sec:derivation}: Derivation of the non-uniform charge and current response coefficients within the independent-particle approximation.
\item Sec.~\ref{sec:time_derivative}: Demonstration of the validity of Eq.~(\ref{eq:CE_coef}) for finite values of the relaxation parameter $\eta$.
\item Sec.~\ref{sec:continuum_model}: Review of the continuum model used to describe $\text{tMoTe}_{\text{2}}$.
\item Sec.~\ref{sec:poisson}: Computational procedures for generating Figs.~\hyperref[fig:patterns]{4(c)} and~\hyperref[fig:patterns]{4(d)}, including the solution of the Poisson equation for the electrostatic potential.
\end{itemize}

\section{\label{sec:derivation}Derivation of Charge and Current Response Coefficients}

In this section, we derive the static charge response coefficient $\zeta^{\alpha_1 \alpha_2}(\mathbf{G}; \omega_0)$ and the static current 
response coefficient $\sigma^{\beta \alpha_1 \alpha_2}(\mathbf{G}; \omega_0)$ within the independent-particle approximation, explicitly accounting for spatial inhomogeneity. 
The derivation proceeds via a perturbative expansion of the reduced density matrix in powers of the external electric field, 
following a methodology similar to that used for computing optical conductivity in the case of uniform photocurrents~\Scite{peres2017,peres2018}.

We work in the length gauge \Scite{peres2017,peres2018}, where the light–matter interaction is described by a dipole coupling term $e\,\mathbf{E}(t) \cdot \mathbf{r}$. 
Here, $e>0$ is the elementary charge, and $\mathbf{E}(t)$ is the time-dependent electric field of the incident light. 
We consider normally incident, monochromatic light with arbitrary polarization:
\begin{equation}
    \mathbf{E}(t) = \mathbf{E}_0(\omega_0) \, e^{-i \omega_0 t} + \mathbf{E}_0(-\omega_0) \, e^{i \omega_0 t},\;\;\; \mathbf{E}_0(-\omega_0) = \mathbf{E}^{*}_0(\omega_0)\nonumber,
\end{equation} 
ensuring $\mathbf{E}(t)$ is real-valued. 

The system Hamiltonian in second-quantized form reads:
\begin{equation}
\label{eq:full_hamiltonian}
    \hat{H}(t) = \sum_{s} \int \left[ \mathrm{d}\mathbf{k} \right] \, \epsilon_{\mathbf{k} s} 
    c^{\dagger}_{\mathbf{k} s} c_{\mathbf{k} s} + i e \mathbf{E}(t) \cdot
    \sum_{s,s^{\prime}} \int \left[ \mathrm{d}\mathbf{k} \right] \,
    c^{\dagger}_{\mathbf{k} s} \mathbf{D}_{\mathbf{k} s s^{\prime}} c_{\mathbf{k} s^{\prime}},
\end{equation}
where
\begin{itemize}
    \item $s,s^{\prime}$ are band indices (including both conduction and valence bands),
    \item $\int \left[ \mathrm{d}\mathbf{k} \right] \equiv \int_{\text{MBZ}} \frac{\mathrm{d}\mathbf{k}}{(2\pi)^2}$ is the integral over the moiré Brillouin zone,
    \item $\epsilon_{\mathbf{k} s}$ is the band dispersion,
    \item $c^{\dagger}_{\mathbf{k} s}$ and $c_{\mathbf{k} s}$ are creation and annihilation operators for Bloch states $\left| \psi_{\mathbf{k} s} \right\rangle$.
\end{itemize}
These operators obey the canonical anticommutation relation:
\begin{equation}
    \{ c_{\mathbf{k} s}, c^{\dagger}_{\mathbf{k}^{\prime} s^{\prime}} \} = (2\pi)^2 \delta_{s s^{\prime}} \delta(\mathbf{k}-\mathbf{k}^{\prime}). \nonumber
\end{equation}
The covariant derivative $\mathbf{D}_{\mathbf{k} s s^{\prime}}$ \Scite{peres2017,peres2018} arises from the position operator in the Bloch representation and is defined as:
\begin{equation}
\label{eq:codrv_def}
    \mathrm{D}^{\beta}_{\mathbf{k} s s^{\prime}} \equiv \delta_{s s^{\prime}} \nabla^{\beta}_{\mathbf{k}} - i \xi^{\beta}_{\mathbf{k} s s^{\prime}}, \;\;\;
    \xi^{\beta}_{\mathbf{k} s s^{\prime}} \equiv i \left\langle u_{\mathbf{k} s} \right| \nabla^{\beta}_{\mathbf{k}} \left| u_{\mathbf{k} s^{\prime}} \right\rangle,
\end{equation} 
where $\xi^{\beta}_{\mathbf{k} s s^{\prime}}$ is the Berry connection. The first term of Eq.~(\ref{eq:full_hamiltonian}) 
describes the unperturbed band structure, while the second term captures the interaction with the optical field.

We focus on the response of the Fourier components of the charge and current density operators at wavevector $\mathbf{G}$ in the moir\'e reciprocal lattice. 
These operators are denoted $\hat{\rho}_{\mathbf{G}}$ and $\hat{\mathbf{j}}_{\mathbf{G}}$, respectively. Their second-quantized forms are given by
\begin{equation}
\label{eq:charge_operator}
    \hat{\rho}_{\mathbf{G}} = -e \sum_{s,s^{\prime}} \int \left[ \mathrm{d}\mathbf{k} \right]\,
    F^{\mathbf{G}}_{\mathbf{k} s s^{\prime}} \, c^{\dagger}_{\mathbf{k} s}  c_{\mathbf{k} s^{\prime}},
\end{equation}

\begin{equation}
\label{eq:current_operator}
    \hat{\mathrm{j}}^{\beta}_{\mathbf{G}} = -e \sum_{s,s^{\prime}} \int \left[ \mathrm{d}\mathbf{k} \right]
     \, \mathrm{M}^{\beta}_{\mathbf{k} s s^{\prime}}(\mathbf{G}) \, c^{\dagger}_{\mathbf{k} s} c_{\mathbf{k} s^{\prime}},
\end{equation}
where $F^{\mathbf{G}}_{\mathbf{k} s s^{\prime}} = \left\langle u_{\mathbf{k} s} \middle| 
u_{\mathbf{k}+\mathbf{G} s^{\prime}} \right\rangle$ is the overlap between the periodic parts of two Bloch states, 
and $\mathrm{M}^{\beta}_{\mathbf{k} s s^{\prime}}(\mathbf{G})$ is defined as \Scite{bradlyn2024uniformity}:
\begin{equation}
    \mathrm{M}^{\beta}_{\mathbf{k} s s^{\prime}}(\mathbf{G}) =  
    \hbar^{-1} \int^1_0 \mathrm{d}\lambda \, \left\langle u_{\mathbf{k} s} \right| 
    [ \nabla^{\beta}_{\mathbf{k}} \mathcal{H}(\mathbf{k}) ]_{\mathbf{k} \to \mathbf{k}+(1-\lambda)\mathbf{G}} \left| u_{\mathbf{k}+\mathbf{G} s^{\prime}} \right\rangle.
\end{equation}
Here, $\mathcal{H}(\mathbf{k}) = e^{-i \mathbf{k} \cdot \hat{\mathbf{r}}} \mathcal{H} e^{i \mathbf{k} \cdot \hat{\mathbf{r}}}$ is the first-quantized single-particle Hamiltonian as a function of $\mathbf{k}$.
As a consistency check, the $\mathbf{G}=\mathbf{0}$ component of Eq.~(\ref{eq:current_operator}) recovers the conventional current operator \Scite{peres2017}:
\begin{equation}
    \hat{\mathrm{j}}^{\beta}_{\mathbf{G}=\mathbf{0}} = -e \sum_{s,s^{\prime}} \int \left[ \mathrm{d}\mathbf{k} \right]
     \, \left\langle u_{\mathbf{k} s} \right| \mathrm{v}^{\beta}(\mathbf{k})
     \left| u_{\mathbf{k} s^{\prime}} \right\rangle 
    \, c^{\dagger}_{\mathbf{k} s} c_{\mathbf{k} s^{\prime}},
\end{equation}
where $\mathrm{v}^{\beta}\left( \mathbf{k} \right) \equiv \hbar^{-1} \nabla^{\beta}_{\mathbf{k}} \mathcal{H}(\mathbf{k})$ is the velocity operator. 

The evolution of the many-body state [described by the density operator $\hat{P}(t)$] under a time-dependent electric field is governed by the Liouville–von Neumann equation:
\begin{equation}
\label{eq:liouville_von_neumann}
    i \hbar \, \frac{\mathrm{d} \hat{P}(t)}{\mathrm{d} t} = [ \hat{H}(t), \hat{P}(t) ].
\end{equation}
To compute the optical response, we focus on the time-dependent expectation values of the Fourier components of the charge and current density operators, 
$\hat{\rho}_{\mathbf{G}}$ and $\hat{\mathbf{j}}_{\mathbf{G}}$ [Eqs.~(\ref{eq:charge_operator}) and~(\ref{eq:current_operator})]. 
Using Eqs.~(\ref{eq:liouville_von_neumann}), we obtain:
\begin{equation}
\label{eq:charge_response}
    \left\langle \hat{\rho}_{\mathbf{G}} \right\rangle (t) \equiv 
    \mathrm{Tr} \left\{ \hat{\rho}_{\mathbf{G}} \hat{P}(t) \right\} 
    = -e  \sum_{s,s^{\prime}} \int \left[ \mathrm{d}\mathbf{k} \right] \,
    F^{\mathbf{G}}_{\mathbf{k} s s^{\prime}} \, P_{\mathbf{k} s^{\prime} s}(t),
\end{equation}
\begin{equation}
\label{eq:current_response}
    \left\langle \hat{\mathrm{j}}^{\beta}_{\mathbf{G}} \right\rangle (t) 
    \equiv \mathrm{Tr} \left\{ \hat{\mathrm{j}}^{\beta}_{\mathbf{G}} \hat{P}(t) \right\}
    = -e \sum_{s,s^{\prime}} \int \left[ \mathrm{d}\mathbf{k} \right] \,
    \mathrm{M}^{\beta}_{\mathbf{k} s s^{\prime}}(\mathbf{G}) 
    \, P_{\mathbf{k} s^{\prime} s}(t),
\end{equation}
where we have introduced the reduced density matrix:
\begin{equation}
\label{eq:RDM_definition}
    P_{\mathbf{k} s s^{\prime}}(t) \equiv \mathrm{Tr} \left\{ c^{\dagger}_{\mathbf{k} s^{\prime}} \, c_{\mathbf{k} s} \hat{P}(t) \right\}.
\end{equation} 
The dynamics of the charge and current responses are thus fully encoded in the time evolution of the reduced density matrix $P_{\mathbf{k}}(t)$. 
To proceed, we derive its equation of motion by differentiating Eq.~(\ref{eq:RDM_definition}) and using Eq.~(\ref{eq:liouville_von_neumann}). This yields:
\begin{equation}
\label{eq:rdm_eom}
    (i \hbar \, \partial_{t} - \Delta \epsilon_{\mathbf{k} s s^{\prime}}) 
    P_{\mathbf{k} s s^{\prime}}(t) = i e \mathbf{E}(t) \cdot [ \mathbf{D}_{\mathbf{k}}, P_{\mathbf{k}}(t) ]_{s s^{\prime}},
\end{equation}
where $\Delta \epsilon_{\mathbf{k} s s^{\prime}} \equiv \epsilon_{\mathbf{k} s} - \epsilon_{\mathbf{k} s^{\prime}}$, and 
$[\mathbf{D}_{\mathbf{k}}, P_{\mathbf{k}} ] = \mathbf{D}_{\mathbf{k}} P_{\mathbf{k}} - P_{\mathbf{k}} \mathbf{D}_{\mathbf{k}}$ denotes the commutator in band space. 
Using the definition of the covariant derivative $\mathbf{D}_{\mathbf{k} s s^{\prime}}$ [Eq.~(\ref{eq:codrv_def})], its commutator with a generic 
operator $\mathcal{O}_{\mathbf{k} s s^{\prime}}$ can be calculated as:
\begin{equation}
\label{eq:codrv_commutator}
    [\mathrm{D}^{\beta}_{\mathbf{k}}, \mathcal{O}_{\mathbf{k}}]_{s s^{\prime}} = \nabla^{\beta}_{\mathbf{k}} 
    \mathcal{O}_{\mathbf{k} s s^{\prime}} - i [\xi^{\beta}_{\mathbf{k}}, \mathcal{O}_{\mathbf{k}}]_{s s^{\prime}}.
\end{equation}
In deriving Eq.~(\ref{eq:rdm_eom}), we have used the decomposition of the Hamiltonian in Eq.~(\ref{eq:full_hamiltonian}), 
separating the unperturbed band structure and the light-matter interaction.

To solve the equation of motion Eq.~(\ref{eq:rdm_eom}), we assume that the system is initially in equilibrium at $t \to -\infty$, 
where the reduced density matrix takes the form $P_{\mathbf{k} s s^{\prime}}(t\to-\infty) = f_{\mathbf{k}s} \delta_{s s^{\prime}}$, with 
$f_{\mathbf{k}s}$ being the Fermi–Dirac distribution. For analytic convenience, we work in frequency space, where Eq.~(\ref{eq:rdm_eom}) becomes:
\begin{equation}
\label{eq:rdm_eom_freq}
    (\hbar\omega-\Delta\epsilon_{\mathbf{k} s s^{\prime}})P_{\mathbf{k} s s^{\prime}}(\omega) = ie \int [ \mathrm{d}\omega^{\prime} ] [ \mathrm{d}\omega^{\prime \prime} ] \, 
    \mathrm{E}^{\beta}(\omega^{\prime}) [ \mathrm{D}^{\beta}_{\mathbf{k}}, P_{\mathbf{k}}(\omega^{\prime \prime})]_{s s^{\prime}} (2\pi)\, \delta(\omega - \omega^{\prime} - \omega^{\prime \prime}).
\end{equation}
Here, $\int [ \mathrm{d}\omega] \equiv \int_{-\infty}^{\infty} \frac{\mathrm{d}\omega}{2\pi}$, and summation over repeated Cartesian indices (e.g., $\beta$) is implied. 
We solve Eq.~(\ref{eq:rdm_eom_freq}) perturbatively, order by order in the electric field, using a recursive approach. 
The $n^{\mathrm{th}}$-order contribution to the reduced density matrix in frequency space is written as
\begin{eqnarray}
\label{eq:nth_RDM}
    P^{(n)}_{\mathbf{k} s s^{\prime}}(\omega) &&= (ie)^n \int \left[ \mathrm{d}\omega_1 \right] 
    ... \left[ \mathrm{d}\omega_n \right] \, \mathrm{E}^{\alpha_1}(\omega_1)...\mathrm{E}^{\alpha_n}(\omega_n) \nonumber\\
    &&\times (2\pi)\, \delta(\omega - \omega_1 - ... -\omega_n) \nonumber\\
    &&\times \mathcal{P}^{\alpha_1 ... \alpha_n}_{\mathbf{k} s s^{\prime}}(\omega_1,...,\omega_n),
\end{eqnarray}
where $\mathcal{P}^{\alpha_1 ... \alpha_n}_{\mathbf{k} s s^{\prime}}(\omega_1,...,\omega_n)$ is a kernel determined recursively. 
The zeroth-order kernel simply corresponds to the equilibrium distribution:
\begin{equation}
\label{eq:0th_kernel}
    \mathcal{P}^{[0]}_{\mathbf{k} s s^{\prime}} = f_{\mathbf{k} s} \delta_{s s^{\prime}},
\end{equation}
where $f_{\mathbf{k} s} = \frac{1}{e^{(\epsilon_{\mathbf{k} s} - \mu)/ (k_B T)} + 1}$. Note that the superscript $[0]$ labels the kernel, 
not the frequency-space density matrix $P^{(n=0)}_{\mathbf{k} s s^{\prime}}(\omega)$. The first-order kernel is given by:
\begin{equation}
\label{eq:1st_kernel}
\mathcal{P}^{\alpha_1}_{\mathbf{k}}(\omega_1) = 
\frac{1}{\hbar \omega_1 - \Delta \epsilon_{\mathbf{k}}} \circ 
[ \mathrm{D}^{\alpha_1}_{\mathbf{k}}, \mathcal{P}^{[0]}_{\mathbf{k}}],
\end{equation}
while the second-order kernel is given by
\begin{eqnarray}
\label{eq:2nd_kernel}
\mathcal{P}^{\alpha_1 \alpha_2}_{\mathbf{k}}(\omega_1, \omega_2) = 
\frac{1}{\hbar \omega_1 + \hbar \omega_2 - \Delta \epsilon_{\mathbf{k}}} \circ 
[ \mathrm{D}^{\alpha_2}_{\mathbf{k}}, \mathcal{P}^{\alpha_1}_{\mathbf{k}}(\omega_1)].
\end{eqnarray}
Here, $\circ$ denotes the Hadamard (element-wise) product of matrices in band indices: $(A\circ B)_{s s^{\prime}} \equiv A_{s s^{\prime}} B_{s s^{\prime}}$.
In this work, we are primarily interested in the second-order response. Specifically, we compute the second-order reduced density matrix as defined by Eq.~(\ref{eq:nth_RDM}) for $n=2$, 
using Eqs.~(\ref{eq:0th_kernel}),~(\ref{eq:1st_kernel}) and~(\ref{eq:2nd_kernel}).
  
Finally, we obtain the nonlinear response coefficients by substituting the second-order reduced density matrix [Eq.~(\ref{eq:nth_RDM}) with $n=2$] into 
the expressions for the charge and current responses, Eqs.~(\ref{eq:charge_response}) and~(\ref{eq:current_response}). 
Retaining only the zero-frequency (DC) component of the response, we derive the second-order charge and current response tensors, 
$\zeta^{\alpha_1 \alpha_2}(\mathbf{G}; \omega_0)$ and $\sigma^{\beta \alpha_1 \alpha_2}(\mathbf{G}; \omega_0)$, respectively:
\begin{eqnarray}
\label{eq:zeta_expression}
    \zeta^{\alpha_1 \alpha_2}(\mathbf{G}; \omega_0) &&= e^3 \int \left[ \mathrm{d}\mathbf{k} \right] \sum_{s,s^{\prime}} 
    F^{\mathbf{G}}_{\mathbf{k} s^{\prime} s} \frac{1}{2i\hbar\eta-\Delta\epsilon_{\mathbf{k} s s^{\prime}}} \left[ \mathrm{D}^{\alpha_2}_{\mathbf{k}},\frac{1}{\hbar\omega_0-\Delta\epsilon_{\mathbf{k}}+i\hbar\eta} 
    \circ[\mathrm{D}^{\alpha_1}_{\mathbf{k}}, \mathcal{P}^{[0]}_{\mathbf{k}}] \right]_{s s^{\prime}} \nonumber\\
    &&+ (\alpha_1,\omega_0 \leftrightarrow \alpha_2,-\omega_0),
\end{eqnarray}
and
\begin{eqnarray}
\label{eq:sigma_expression}
    \sigma^{\beta \alpha_1 \alpha_2}(\mathbf{G}; \omega_0) &&= e^3 \int \left[ \mathrm{d}\mathbf{k} \right] \sum_{s,s^{\prime}} 
    M^{\beta}_{\mathbf{k} s^{\prime} s}(\mathbf{G}) \frac{1}{2i\hbar\eta-\Delta\epsilon_{\mathbf{k} s s^{\prime}}} \left[ \mathrm{D}^{\alpha_2}_{\mathbf{k}},\frac{1}{\hbar\omega_0-\Delta\epsilon_{\mathbf{k}}+i\hbar\eta} 
    \circ[\mathrm{D}^{\alpha_1}_{\mathbf{k}}, \mathcal{P}^{[0]}_{\mathbf{k}}] \right]_{s s^{\prime}} \nonumber\\
    &&+ (\alpha_1,\omega_0 \leftrightarrow \alpha_2,-\omega_0).
\end{eqnarray}
Here, $\eta=0^+$ is an infinitesimal positive parameter introduced to regularize the frequency denominators. 
Physically, this corresponds to adiabatically switching on the electric field in the distant past:
\begin{equation}
    \mathbf{E}(t) = e^{\eta t} \left[ \mathbf{E}_0(\omega_0) \, e^{-i \omega_0 t} + \mathbf{E}_0(-\omega_0) \, e^{i \omega_0 t} \right].
\end{equation} 
The notation $(\alpha_1,\omega_0 \leftrightarrow \alpha_2,-\omega_0)$ indicates symmetrization under the exchange of indices and frequencies, ensuring that the response functions satisfy
\begin{equation}
    \zeta^{\alpha_1 \alpha_2}(\mathbf{G}; \omega_0) = \zeta^{\alpha_2 \alpha_1}(\mathbf{G}; -\omega_0),\;\;\; 
\sigma^{\beta \alpha_1 \alpha_2}(\mathbf{G}; \omega_0) = \sigma^{\beta \alpha_2 \alpha_1}(\mathbf{G}; -\omega_0).
\end{equation}

From Eq.~(\ref{eq:zeta_expression}), we observe that the terms with $s=s^{\prime}$ in the integrand contain a prefactor $\frac{1}{2i\hbar\eta}$, 
indicating a divergence of order $\mathcal{O}(\eta^{-1})$. To isolate this divergent contribution, we simplify the nested commutator in 
Eq.~(\ref{eq:zeta_expression}) using Eq.~(\ref{eq:codrv_commutator}), assuming a cold semiconductor for simplicity. This yields:
\begin{eqnarray}
\label{eq:commutator_simplify}
    &&\left[ \mathrm{D}^{\alpha_2}_{\mathbf{k}},\frac{1}{\hbar\omega_0-\Delta\epsilon_{\mathbf{k}}+i\hbar\eta} \circ[\mathrm{D}^{\alpha_1}_{\mathbf{k}}, \mathcal{P}^{[0]}_{\mathbf{k}}] \right]_{s s^{\prime}} \nonumber \\
    =&& \nabla^{\alpha_2}_{\mathbf{k}} \left( \frac{-i \xi^{\alpha_1}_{\mathbf{k} s s^{\prime}} \Delta f_{\mathbf{k} s^{\prime} s}}{\hbar \omega_0 - \Delta \epsilon_{\mathbf{k} s s^{\prime}} + i\hbar\eta} \right) 
    - \sum_r \left( \frac{\xi^{\alpha_1}_{\mathbf{k} r s^{\prime}} \xi^{\alpha_2}_{\mathbf{k} s r} \Delta f_{\mathbf{k} s^{\prime} r}}{\hbar \omega_0 - \Delta \epsilon_{\mathbf{k} r s^{\prime}} + i\hbar\eta} - 
    \frac{\xi^{\alpha_1}_{\mathbf{k} s r} \xi^{\alpha_2}_{\mathbf{k} r s^{\prime}} \Delta f_{\mathbf{k} r s}}{\hbar \omega_0 - \Delta \epsilon_{\mathbf{k} s r} + i\hbar\eta} \right).
\end{eqnarray}
Substituting Eq.~(\ref{eq:commutator_simplify}) into Eq.~(\ref{eq:zeta_expression}) and focusing on the $s=s^{\prime}$ terms, the response coefficient becomes:
\begin{equation}
    \zeta^{\alpha_1 \alpha_2}(\mathbf{G}; \omega_0) \to \frac{i e^3}{2\hbar\eta} \int \left[ \mathrm{d}\mathbf{k} \right] \sum_{r \ne s} \left( F^{\mathbf{G}}_{\mathbf{k} ss} - F^{\mathbf{G}}_{\mathbf{k} rr} \right)
    \frac{\xi^{\alpha_1}_{\mathbf{k} r s} \xi^{\alpha_2}_{\mathbf{k} s r} \Delta f_{\mathbf{k} s r}}{\hbar \omega_0 - \Delta \epsilon_{\mathbf{k} r s} + i\hbar\eta}
    + (\alpha_1,\omega_0 \leftrightarrow \alpha_2,-\omega_0).
\end{equation}
To extract the $\mathcal{O}(\eta^{-1})$ divergent part, we use the identity \Scite{gao2021intrinsic}:
\begin{equation}
    \frac{1}{\eta} \frac{1}{x+i \eta} = \frac{1}{\eta} \left[ \frac{\mathcal{P}}{x} - i \pi \delta(x) \right] - i\frac{\mathcal{P}}{x^2}, 
\end{equation}
where $\mathcal{P}$ denotes the Cauchy principal value. Keeping only the leading $\mathcal{O}(\eta^{-1})$ divergent contribution, we obtain the so-called accumulation response:
\begin{equation}
\label{eq:zeta_inj_expression}
    \zeta_{\mathrm{accum}}^{\alpha_1 \alpha_2}(\mathbf{G}; \omega_0) = \frac{\pi e^3}{\hbar\eta} 
    \int \left[ \mathrm{d}\mathbf{k} \right] \sum_{r \ne s} \left( F^{\mathbf{G}}_{\mathbf{k} ss} - F^{\mathbf{G}}_{\mathbf{k} rr} \right) 
    \, \xi^{\alpha_1}_{\mathbf{k} rs} \xi^{\alpha_2}_{\mathbf{k} sr} \Delta f_{\mathbf{k} sr} \delta(\hbar\omega_0 - \Delta\epsilon_{\mathbf{k} rs}).
\end{equation}
This expression corresponds precisely to Eq.~(\ref{eq:zeta_inj}) in the main text.

\section{\label{sec:time_derivative}Derivation of Eq.~(\ref{eq:CE_coef})}

As noted in the main text [see footnote above Eq.~(\ref{eq:CE_coef})], the identity 
\begin{equation}
\label{eq:CE_coef_sm}
    2\eta\,\zeta^{\alpha_1\alpha_2}(\mathbf{G};\omega_0) + i \mathrm{G}^{\beta} \sigma^{\beta \alpha_1\alpha_2}(\mathbf{G};\omega_0) = 0
\end{equation}
remains valid for finite $\eta$ if $\zeta^{\alpha_1\alpha_2}(\mathbf{G};\omega_0)$ is interpreted as the full response 
[Eq.~(\ref{eq:zeta})], not just the accumulation part [Eq.~(\ref{eq:zeta_inj})]. We derive Eq.~(\ref{eq:CE_coef_sm}) below 
to justify the generalized form of Eq.~(\ref{eq:CE_coef}), which is useful when $\eta$ is taken to be small but finite 
in our numerical calculations. This result follows directly from the continuity equation. At the operator level, the 
continuity equation reads:
\begin{equation}
\label{eq:continuity_equation_operator}
    \partial_t \hat{\rho}_{\mathbf{G}} + i \mathbf{G} \cdot \hat{\mathbf{j}}_{\mathbf{G}} = 0.
\end{equation}
In the Schr\"{o}dinger picture, operators such as $\hat{\rho}_{\mathbf{G}}$ [as defined in Eq.~(\ref{eq:charge_operator})] are time-independent, so
$\partial_t \hat{\rho}_{\mathbf{G}} = 0$ identically. Therefore, Eq.~(\ref{eq:continuity_equation_operator}) must be interpreted in the Heisenberg picture, 
where operators evolve with time:
\begin{equation}
\label{eq:CE_H}
    \partial_t \hat{\rho}^{(\mathrm{H})}_{\mathbf{G}} + i \mathbf{G} \cdot \hat{\mathbf{j}}^{(\mathrm{H})}_{\mathbf{G}} = 0.
\end{equation}
Here, the superscript $(\mathrm{H})$ indicates operators in the Heisenberg picture; operators without it are understood to be in the Schr\"{o}dinger picture. 
Taking the expectation value of both sides of Eq.~(\ref{eq:CE_H}), we obtain:
\begin{equation}
\label{eq:CE_expectation}
    \left\langle  \partial_t \hat{\rho}^{(\mathrm{H})}_{\mathbf{G}} \right\rangle + 
    i \mathbf{G} \cdot \left\langle  \hat{\mathbf{j}}^{(\mathrm{H})}_{\mathbf{G}} \right\rangle = 0.
\end{equation}

To evaluate the expectation value $\left\langle \partial_t \hat{\rho}^{(\mathrm{H})}_{\mathbf{G}} \right\rangle$, we proceed as follows:
\begin{eqnarray}
\label{eq:trace_trick}
    \left\langle \partial_t \hat{\rho}^{(\mathrm{H})}_{\mathbf{G}} \right\rangle && \equiv \mathrm{Tr} 
    \left\{ \partial_t \hat{\rho}^{(\mathrm{H})}_{\mathbf{G}} \hat{P}^{(\mathrm{H})}(t) \right\} \nonumber\\
    &&= \frac{1}{i \hbar} \mathrm{Tr} \left\{ [\hat{\rho}^{(\mathrm{H})}_{\mathbf{G}}, \hat{H}^{(\mathrm{H})}] \hat{P}^{(\mathrm{H})}(t) \right\} \nonumber\\
    &&= \frac{1}{i \hbar} \mathrm{Tr} \left\{ \hat{\rho}^{(\mathrm{H})}_{\mathbf{G}} [\hat{H}^{(\mathrm{H})}, \hat{P}^{(\mathrm{H})}(t)] \right\} \nonumber\\
    &&= \frac{1}{i \hbar} \mathrm{Tr} \left\{ \hat{\rho}_{\mathbf{G}} [\hat{H}, \hat{P}(t)] \right\} \nonumber\\
    &&= \mathrm{Tr} \left\{ \hat{\rho}_{\mathbf{G}} \frac{\mathrm{d}}{\mathrm{d} t} \hat{P}(t) \right\},
\end{eqnarray}
where:
\begin{itemize}
    \item The second line follows from the Heisenberg equation of motion.
    \item The third line uses the cyclic property of the trace: $\mathrm{Tr}\{[A,B]C\} = \mathrm{Tr}\{A[B,C]\}$.
    \item The fourth line switches back to the Schr\"{o}dinger picture, since the trace is invariant under picture change.
    \item The final line applies the Liouville–von Neumann equation [Eq.~(\ref{eq:liouville_von_neumann})].
\end{itemize}
We now focus on the second-order contribution to the density operator, $\hat{P}^{(2)}(t)$. This component is related to its frequency-domain representation by: 
$\hat{P}^{(2)}(t) = \int \left[ \mathrm{d}\omega \right] e^{-i\omega t + 2 \eta t} \hat{P}^{(2)}(\omega)$. 
Taking the time derivative in Eq.~(\ref{eq:trace_trick}) and setting $\omega=0$ then yields 
\footnote{The precise form of the time derivative of $\hat{P}^{(2)}(t)$ depends on how the relaxation parameter $\uptau$ is introduced. As a result, the 
coefficient $2\eta$ in Eq.~(\ref{eq:ptrho_sm}) may not be universal.}:
\begin{equation}
\label{eq:ptrho_sm}
    \left\langle \partial_t \hat{\rho}^{(\mathrm{H})}_{\mathbf{G}} \right\rangle = 
    2\eta\,\mathrm{Tr} \left\{ \hat{\rho}_{\mathbf{G}} \hat{P}^{(2)}(t) \right\} =  
    2 \eta \, \Delta\rho_{\mathbf{G}}.
\end{equation}
Substituting this result into Eq.~(\ref{eq:CE_expectation}), we obtain:
\begin{equation}
    2 \eta \, \Delta\rho_{\mathbf{G}} + i \mathbf{G} \cdot \left\langle \hat{\mathbf{j}}^{(\mathrm{H})}_{\mathbf{G}} \right\rangle = 0.
\end{equation}
Finally, expressing $\Delta\rho_{\mathbf{G}}$ and $\left\langle \hat{\mathbf{j}}^{(\mathrm{H})}_{\mathbf{G}} \right\rangle$ in terms of 
the response coefficients $\zeta^{\alpha_1\alpha_2}(\mathbf{G};\omega_0)$ and $\sigma^{\beta \alpha_1\alpha_2}(\mathbf{G};\omega_0)$, 
we recover Eq.~(\ref{eq:CE_coef_sm}).

\section{\label{sec:continuum_model}Continuum Model Hamiltonian of $\text{tMoTe}_{\text{2}}$}

The continuum model Hamiltonian of $\text{tMoTe}_{\text{2}}$ at twist angle $\theta$, for the $+\mathbf{K}$ valley and spin-up electrons, is given by the 
following $4\times 4$ matrix \Scite{wu2019topological}:
\begin{equation}
\mathcal{H}_{\uparrow}(\mathbf{k}, \mathbf{r})=\left(\begin{array}{cc}
h_b(\mathbf{k}, \mathbf{r}) & T(\mathbf{r}) \\
T^{\dagger}(\mathbf{r})     & h_t(\mathbf{k}, \mathbf{r})
\end{array}\right),
\end{equation}
where $h_l(\mathbf{k}, \mathbf{r})$ (with $l=+1$ for the bottom layer and $l=-1$ for the top layer) describes the intralayer Hamiltonian, and $T(\mathbf{r})$ encodes the interlayer coupling. Here, $\mathbf{k} \equiv -i\nabla_{\mathbf{r}}$ acts as an operator on the real-space coordinate $\mathbf{r}$. The 
corresponding Hamiltonian for the $-\mathbf{K}$ valley and spin-down electrons can be obtained via time-reversal symmetry.

The intralayer Hamiltonians take the form:
\begin{equation}
h_l(\mathbf{k}, \mathbf{r}) = \left(\begin{array}{cc}
\Delta_g + \Delta_{l,c}(\mathbf{r}) & 0 \\
0                                   & \Delta_{l,v}(\mathbf{r}) 
\end{array}\right) + e^{-i l \frac{\theta}{4} \xi_z} \left[ \hbar v_F (\mathbf{k} - \boldsymbol{\kappa}_l) \cdot \boldsymbol{\xi} \right] e^{i l \frac{\theta}{4} \xi_z}
,
\end{equation}
where the subscripts $c$ and $v$ label the conduction and valence bands of the monolayer, respectively. The vector $\boldsymbol{\xi}$ contains the Pauli matrices acting 
in the band subspace, and $\boldsymbol{\kappa}_l$ denotes the $+\mathbf{K}$ point of monolayer $\text{MoTe}_{\text{2}}$ in layer $l$, as defiend in 
Fig.~\hyperref[fig:tMoTe2]{2(c)} of the main text. 
The spatially modulated potentials $\Delta_{l,\nu}(\mathbf{r})$ (with $\nu=c,\,v$) are given by:
\begin{equation}
    \Delta_{l,\nu}(\mathbf{r}) = 2 V_{\nu} \sum_{i=1,3,5} \mathrm{cos}(\mathbf{G}_i \cdot \mathbf{r} + l \psi_{\nu}),
\end{equation}
with $\mathbf{G}_i$ denoting the moir\'e reciprocal lattice vectors defined in Fig.~\hyperref[fig:tMoTe2]{2(c)} of the main text.

The interlayer coupling matrix $T(\mathbf{r})$ is expressed as:
\begin{equation}
    T(\mathbf{r}) = T_0 + T_1 e^{i \mathbf{G}_5 \cdot \mathbf{r}} + T_2 e^{i \mathbf{G}_6 \cdot \mathbf{r}},
\end{equation}
with the matrices $T_0$, $T_1$, $T_2$ defined as:
\begin{equation}
T_0 = \left(\begin{array}{cc}
w_{c}  & w_{cv} \\
w_{vc} & w_{v} 
\end{array}\right),\;\;\;
T_1 = \left(\begin{array}{cc}
w_{c}              & w_{cv} e^{-i2\pi/3} \\
w_{vc} e^{i2\pi/3} & w_{v} 
\end{array}\right),\;\;\;
T_2 = \left(\begin{array}{cc}
w_{c}               & w_{cv} e^{i2\pi/3} \\
w_{vc} e^{-i2\pi/3} & w_{v} 
\end{array}\right).
\end{equation}
The parameter values used in this model are:
\begin{itemize}
    \item $\Delta_g=1.1\, \text{eV}$,
    \item $v_F = 0.4\times10^6 \, \text{m/s}$,
    \item $(V_v, \psi_v, w_v) = (8\, \text{meV}, -89.6^\circ, -8.5\, \text{meV})$,
    \item $(V_c, \psi_c, w_c) = (5.97\, \text{meV}, -87.9^\circ, -2\, \text{meV})$,
    \item $w_{cv} = w_{vc}^* = 15.3\, \text{meV}$.
\end{itemize}
The monolayer lattice constant of $\text{MoTe}_{\text{2}}$ is taken to be $3.472$~\AA.

\section{\label{sec:poisson}Technical Details for Generating Figs.~\hyperref[fig:patterns]{4(c)} and~\hyperref[fig:patterns]{4(d)}}

In this section, we outline the procedure used to compute the electrostatic potential $U(\mathbf{r})$ generated by the redistributed charge in the $\text{tMoTe}_{\text{2}}$ system, which is used to produce Figs.~\hyperref[fig:patterns]{4(c)} and~\hyperref[fig:patterns]{4(d)}. 

The system is modeled in a Cartesian coordinate system. The top layer of $\text{MoTe}_{\text{2}}$, located at $z=0$, 
carries a surface charge density $\Delta\rho_{\mathrm{2D}}^t(x,y)$, while the bottom layer, located at $z=-d_0$, carries a surface charge density 
$\Delta\rho_{\mathrm{2D}}^b(x,y)$, where $d_0$ is the interlayer separation. An adjacent layered material is placed at $z=d_1>0$, and our objective is to 
determine the electrostatic potential $U(x,y,z=d_1)$. For the numerical results shown in the main text, we use $d_0 = 6.9$~\AA~and $d_1 = 10$~\AA.

The electrostatic potential $U(x,y,z)$ satisfies the Poisson equation:
\begin{equation}
\label{eq:poisson}
    \nabla^2 U(x,y,z) = -\frac{1}{\varepsilon_0}\left[ \Delta\rho_{\mathrm{2D}}^t(x,y) \delta(z) + \Delta\rho_{\mathrm{2D}}^b(x,y) \delta(z+d_0) \right],
\end{equation}
where $\varepsilon_0$ is the vacuum permittivity, and $\delta(z)$ represents the Dirac delta function. To compute the source term on the right-hand side 
of Eq.~(\ref{eq:poisson}), which includes $\Delta\rho_{\mathrm{2D}}^t(x,y)$ and $\Delta\rho_{\mathrm{2D}}^b(x,y)$, we evaluate the layer-resolved charge response 
coefficients $\zeta^{\alpha_1 \alpha_2}(\mathbf{G}, l; \omega_0)$. These coefficients are obtained 
by replacing the original wavefunctions $\left| u_{\mathbf{k} s} \right\rangle$ with the layer-projected wavefunctions 
$\left| u^{l}_{\mathbf{k} s} \right\rangle$ in the form factor $F^{\mathbf{G}}_{\mathbf{k} s s^{\prime}}$ in Eq.~(\ref{eq:zeta_expression}).

To solve Eq.~(\ref{eq:poisson}), we apply a Fourier transform:
\begin{equation}
\label{eq:poisson_fourier}
    U(x,y,z) = \sum_{\mathbf{G}} \int \frac{\mathrm{d}k}{2\pi}\, u_\mathbf{G}(k) \, e^{i\mathbf{G}\cdot \mathbf{r}_{\mathrm{2D}}+ikz},
\end{equation}
where the sum is over moir\'e reciprocal lattice vectors $\mathbf{G}$, and $\mathbf{r}_{\mathrm{2D}}=(x,y)$. Substituting Eq.~(\ref{eq:poisson_fourier}) 
into Eq.~(\ref{eq:poisson}) transforms the differential equation into an algebraic equation for $u_\mathbf{G}(k)$, which can be easily solved:
\begin{equation}
\label{eq:fourier_sol}
    u_\mathbf{G}(k) = \frac{\Delta\rho_{\mathbf{G}}^t + \Delta\rho_{\mathbf{G}}^b \, e^{ikd_0}}{\varepsilon_0 \,(|\mathbf{G}|^2+k^2)}.
\end{equation}
Inserting Eq.~(\ref{eq:fourier_sol}) into Eq.~(\ref{eq:poisson_fourier}) and performing the $k$-integral yields the electrostatic potential at $z=d_1$:
\begin{equation}
\label{eq:poisson_sol}
    U(x,y,z=d_1) = \sum_{\mathbf{G}} u_{\mathbf{G}} \, e^{i\mathbf{G}\cdot \mathbf{r}_{\mathrm{2D}}},
\end{equation}
with coefficients
\begin{equation}
\label{eq:potential_fourier}
    u_{\mathbf{G}} = \frac{e^{-|\mathbf{G}| d_1}}{2 \varepsilon_0 |\mathbf{G}|} (\Delta\rho_{\mathbf{G}}^t + \Delta\rho_{\mathbf{G}}^b \, e^{-|\mathbf{G}| d_0}).
\end{equation}
The numerical results obtained from Eqs.~(\ref{eq:poisson_sol}) and~(\ref{eq:potential_fourier}) are then used to generate the electrostatic potential 
distributions shown in Figs.~\hyperref[fig:patterns]{4(c)} and~\hyperref[fig:patterns]{4(d)} of the main text.

\bibliography{ref}